\documentclass{elsart}
\usepackage{graphicx}
\def\be{\begin{equation}}
\def\ee{\end{equation}}
\def\bea{\begin{eqnarray}}
\def\eea{\end{eqnarray}}
\def\beq{\begin{equation}}
\def\eeq{\end{equation}}
\def\beqa{\begin{eqnarray}}
\def\eeqa{\end{eqnarray}}
\def\bce{\begin{center}}
\def\ece{\end{center}}

\def\<{\langle}
\def\>{\rangle}

\newcommand{\AmS}{{\protect\the\textfont2
  A\kern-.1667em\lower.5ex\hbox{M}\kern-.125emS}}

\newcommand{\nn}{\nonumber}
\newcommand{\nnl}{\nonumber\\}

\newcommand{\hdag}{^\dagger}
\newcommand{\hdagb}{^\dagger}
\newcommand{\sign}{\Sigma_c^{0}}
\newcommand{\sigp}{\Sigma_c^{+}}
\newcommand{\sigpp}{\Sigma_c^{++}}

\newcommand{\signpp}{\Sigma_c^{++,0}}

\newcommand{\matel}[3]{\langle\,#1\,|\,#2\,|\,#3\,\rangle}
\newcommand{\matelsym}[2]{\matel{#1}{#2}{#1}}
\newcommand{\wf}[1]{|\,#1\,\rangle}
\newcommand{\wfs}[1]{|#1\;\rangle}
\newcommand{\wfbig}[1]{\left|\,#1\,\right\rangle}
\newcommand{\Cms}{\chi_{ms}}
\newcommand{\Cma}{\chi_{ma}}
\newcommand{\Pms}{\phi_{ms}}
\newcommand{\Pma}{\phi_{ma}}
\newcommand{\protms}{uud+udu-2duu}
\newcommand{\neutms}{ddu+dud-2udd}
\newcommand{\protma}{uud-udu}
\newcommand{\neutma}{dud-ddu}
\newcommand{\ccums}{ccu+cuc-2ucc}
\newcommand{\ccdms}{ccd+cdc-2dcc}
\newcommand{\ccuma}{cuc-ccu}
\newcommand{\ccdma}{cdc-ccd}
\newcommand{\uscms}{(ucs+scu)+(usc+suc)-2(csu+cus)}
\newcommand{\uscma}{(usc+suc)-(scu+ucs)}
\newcommand{\dscms}{(scd+dcs)+(sdc+dsc)-2(cds+csd)}
\newcommand{\dscma}{(sdc+dsc)-(dcs+scd)}
\newcommand{\scpms}{(ucd+dcu)+(udc+duc)-2(cdu+cud)}
\newcommand{\scpma}{(udc+duc)-(dcu+ucd)}

\newcommand{\lcpms}{(udc-duc)+(ucd-dcu)}
\newcommand{\lcpma}{(ucd-dcu)+(duc-udc)-2(cdu-cud)}

\newcommand{\scnms}{ddc+dcd-2cdd}
\newcommand{\scnma}{ddc-dcd}
\newcommand{\scppms}{uuc+ucu-2cuu}
\newcommand{\scppma}{uuc-ucu}
\newcommand{\spupms}{\uparrow\uparrow\downarrow+\uparrow\downarrow
\uparrow-2\downarrow\uparrow\uparrow}
\newcommand{\spupma}{\uparrow\uparrow\downarrow-\uparrow\downarrow\uparrow}
\newcommand{\spdnms}{\downarrow\downarrow\uparrow+\downarrow\uparrow
\downarrow-2\uparrow\downarrow\downarrow}
\newcommand{\spdnma}{\downarrow\uparrow\downarrow-\downarrow\downarrow\uparrow}

\begin{document}
\begin{frontmatter}
\title{Bound States of Heavy Flavor Hyperons}

\author[ugi,hip]{F. Fr\"omel,}
\author[hip]{B. Juli\'a-D\'{\i}az}
\and 
\author[hip]{D. O. Riska}

\address[ugi]{Institut f\"ur Theoretische Physik, 
	Universit\"at Giessen,  \\
	Heinrich-Buff-Ring 16, 
	D-35392 Giessen, Germany }

\address[hip]{Helsinki Institute of Physics and
	  Department of Physical Sciences, \\
        POB 64, 00014 University of Helsinki, Finland}

\begin{abstract}
Several realistic phenomenological nucleon-nucleon interaction models are 
employed to investigate the possibility of bound deuteron-like
states of such heavy flavor hyperons and nucleons, for which the
interaction between the light flavor quark components is expected
to be the most significant interaction. The results indicate that
deuteron-like bound states are likely
to form between nucleons and the $\Xi_c^{'}$ and $\Xi_{cc}$ charm 
hyperons as well as between $\Xi$ hyperons and double-charm 
hyperons. Bound states between two $\Sigma_c$ hyperons
are also likely. In the case of beauty hyperons the corresponding
states are likely to be deeply bound.

\end{abstract}
\end{frontmatter}

\section{Introduction}

The interaction between nucleons and several classes of heavy
flavor hyperons is expected to be dominated by the long range
interaction between the light flavor components of the baryons.
As an example, the interaction between nucleons and the
recently discovered double-charm hyperons~\cite{SELEX1,SELEX2,SELEX3} 
is mainly due to the interaction between the single light flavor quark
in the
double flavor hyperon and those in the nucleon. The color-neutral
interaction
between charm and light flavor quarks is either weak or of
very short range. In the case of two-baryon
states, in which the interaction between the light flavor
quarks is the dominant one, it should be possible to calculate
the binding energy to a first approximation
by modifying realistic phenomenological
nucleon-nucleon interaction models
to take into account the different numbers
of light flavor quarks in the baryons~\cite{bruno}.
This method is explored here. It is found to have at
most qualitative value, however, due to the fact that 
the short range components of the extant
realistic phenomenological nucleon-nucleon 
potentials are poorly constrained. Its use should 
be limited to those potentials which do not 
have strong angular momentum dependence.

The modification of the nucleon-nucleon interaction
required for systems with different numbers of light
flavor quarks is straightforward when the nucleon-nucleon 
interaction is expressed in terms of operators,
which have well defined matrix elements in the quark model,
and the strength of which may therefore be 
correspondingly rescaled from the nucleon-nucleon
system
to the two-baryon system under consideration. In this
approach the radial behavior of the interaction
components is determined by the phenomenological
nucleon-nucleon interaction, without any need for a
detailed microscopic quark model based derivation of the
interaction. Two-baryon systems that may
--- at least approximately --- be described in this way are
the following: $N-\Sigma_c$, $N-\Xi_c^{'}$, $N-\Xi_{cc}$ as
well as those with the nucleon ($N$) replaced by the
paired charm hyperon and finally also the corresponding
states that involve beauty rather than charm hyperons.
To the extent that the interaction between the strange
and light flavor quarks may be neglected the
interaction between $\Xi$ hyperons and
the corresponding charm hyperons may also be
described in this way in a first approximation. In contrast 
the interactions of heavy flavor hyperons with zero isospin 
like the $\Lambda_c$ and the $\Lambda_b$ hyperons,
which do
not couple to pions, cannot be approximated by rescaled
versions of the nucleon-nucleon interaction.

The interaction between nucleons and charm and double-charm
hyperons is weaker than that between the nucleons, because
of the smaller number of light flavor quarks in the charm
and double-charm hyperons. The weaker attractive interaction
is largely compensated, however, by the weakening of the
repulsive effect of the two-baryon kinetic energy that is
caused by the larger masses of the charm and double-charm
hyperons. As a consequence the calculated binding energies
remain small and dependent on the details of the interaction
model in the case of charm hyperons, and typically
become large only in the case of systems of beauty
hyperons.

Here the interaction between the two-baryon states that
are formed of nucleons and heavy flavor hyperons is
calculated from the realistic phenomenological 
interaction models in Refs.~\cite{nijm,AV18}.
For an estimate of the theoretical uncertainty
in the calculated binding energies the set ``AVn'' 
of systematically simplified versions of the AV18
interaction model is also employed~\cite{pieper}.   

Section 2 below contains a description 
two-baryon states that are formed of nucleons 
and iso-doublet hyperons. Two-baryon states with isospin 
1 hyperons are considered in section 3. The results are 
summarized in the concluding discussion.

\section{Two-baryon states of isospin 1/2 baryons}
\label{sec:isohalf}

Two-particle states formed of heavy flavor
isodoublet baryons are similar to
the two-nucleon system in that they 
have a long range pion exchange interaction, and an attractive
intermediate range component, which to a large extent
may be attributed to two-pion exchange~\cite{chemtob,bira}.
To the extent that the color-neutral interactions between their heavy
flavor quarks
is weak in comparison to the interaction between the
light flavor quark components, the interaction may ---
to a first approximation --- be constructed by multiplication 
of the components of the nucleon-nucleon interaction 
by appropriate quark model scaling factors. 

The phenomenological nucleon-nucleon interaction is in general
expressible in terms of rotational 
invariants of spin and isospin operators as well as 
momenta and angular momenta. The scaling factors for the
strengths of the matrix elements of spin and isospin
invariants for two baryon states of isodoublet baryons
relative to the corresponding two-nucleon state matrix
elements may be derived from the quark model matrix
elements of the spin and isospin operators for light
flavor quarks. For $\Xi_{cc}$ (and $\Xi$) hyperon
states the matrix elements are:
\bea
	&&\matelsym{\Xi_{(cc)}}{\sum_q 1^q}      
		= \frac{1}{3}\matelsym{N}{\sum_q 1^q}\, ,
\nonumber\\ 
	&&\matelsym{\Xi_{(cc)}}{\sum_q\sigma_i^q}  
		=-\frac{1}{3}\matelsym{N}{\sum_q\sigma_i^q}\, , 
\nonumber\\ 
	&&\matelsym{\Xi_{(cc)}}{\sum_q\tau_i^q}    
		= \matelsym{N}{\sum_q\tau_i^q}\, ,
\nonumber\\ 
	&&\matelsym{\Xi_{(cc)}}{\sum_q\sigma_i^q\tau_j^q} 
		= -\frac{1}{5}\matelsym{N}{\sum_q\sigma_i^q\tau_j^q}
\, . 
\eea
For the corresponding single charm hyperon states formed
of $\Xi_c^{'}$ hyperons the matrix elements are:
\bea
	&&\matelsym{\Xi_c^{'}}{\sum_q 1^q}      
		= \frac{1}{3}\matelsym{N}{\sum_q 1^q}\ , \nnl
	&&\matelsym{\Xi_c^{'}}{\sum_q\sigma_i^q}  
		= \frac{2}{3}\matelsym{N}{\sum_q\sigma_i^q}\, , \nnl
	&&\matelsym{\Xi_c^{'}}{\sum_q\tau_i^q}    
		= \matelsym{N}{\sum_q\tau_i^q} \, ,  \nnl
	&&\matelsym{\Xi_c^{'}}{\sum_q\sigma_i^q\tau_j^q} 
		= \frac{2}{5}\matelsym{N}{\sum_q\sigma_i^q\tau_j^q}
\, .
\eea

 From these matrix elements one may derive the scaling
relations that apply for the rotational operators in nucleon-nucleon
interactions. As an example the AV18 interaction
model is given in the operator form~\cite{AV18}:

\bea
&&V=\sum_{p=1}^{18}v_p(r) O^p\, ;\\
&&O^{p=1,..14}=[1,\vec\sigma_i\cdot\vec\sigma_j,
S_{ij},\vec L\cdot \vec S,\vec L^2, 
\vec L^2 \vec\sigma_i\cdot\vec\sigma_j, 
(L\cdot S)^2]\otimes[1,\vec\tau_i
\cdot \vec \tau_j]\, .
\label{opv}
\eea
The operators $O^{p=15,..18}$ are part of the isospin breaking
electromagnetic components which depend on the baryon charge. 
These (small) terms are dropped here since they are inapplicable 
to hyperons.

The decomposition used in Ref.~\cite{nijm} is slightly different
from (\ref{opv}):
\bea
&&O^{p=1,..10}=[1,\vec\sigma_i\cdot\vec\sigma_j,
S_{ij},\vec L\cdot \vec S,Q_{ij}]
\otimes[1,\vec\tau_i\cdot \vec \tau_j]\, ,
\label{opv2}	
\eea
where $Q_{ij}=\frac{1}{2}[(\vec\sigma_i\cdot \vec L)
(\vec\sigma_j\cdot \vec L)+(\vec\sigma_j\cdot \vec L)
(\vec\sigma_i\cdot \vec L)]$.
The corresponding scaling factors are 
listed in Table~\ref{tab:scl_xi}. 

\begin{table}
  \begin{center}
  \begin{tabular}{c|cc|cc|c}
  \hline\hline
  			&$\Xi_{(cc)}-N$	&$\Xi_{(cc)}-\Xi_{(cc)}$

& $\Xi_c^{'}-N$ & $\Xi_c^{'}-\Xi_c^{'}$
									
		& $\Xi_c^{'}-\Xi_{(cc)}$ \\
  \hline
  Operator		& \multicolumn{5}{c}{Scaling factor}
						\\
  \hline\hline 
  $1$			& $1/3$		& $1/9$		& $1/3$	
	& $1/9$		& $1/9$		\\
  $\vec\tau_i\cdot\vec\tau_j$
			& $1$		& $1$		& $1$	
	& $1$		& $1$		\\
  $\vec\sigma_i\cdot\vec\sigma_j$ 
			& $-1/3$	& $1/9$		& $2/3$	
	& $4/9$		& $-2/9$	\\
  $(\vec\sigma_i\cdot\vec\sigma_j)(\vec\tau_i\cdot\vec\tau_j)$
			& $-1/5$	& $1/25$	& $2/5$	
	& $4/25$	& $-2/25$	\\
  $S_{ij}$		
			& $-1/3$	& $1/9$		& $2/3$	
	& $4/9$		& $-2/9$	\\
  $S_{ij}(\vec\tau_i\cdot\vec\tau_j)$
			& $-1/5$	& $1/25$	& $2/5$	
	& $4/25$	& $-2/25$	\\

  $\vec L\cdot\vec S$	
			& $1/3\hdagb$	& $-1/9$	& $2/3\hdagb$
	& $2/9$		& $2/9\hdagb$	\\
  $\vec L\cdot\vec S (\vec\tau_i\cdot\vec\tau_j)$
			& $1\hdagb$	& $-1/5$	& $1\hdagb$
	& $2/5$		& $2/5\hdagb$ \\
\hline
  $L^2$			
			& $1/3$		& $1/9$		& $1/3$	
	& $1/9$		& $1/9$		\\
  $L^2(\vec\tau_i\cdot\vec\tau_j)$
			& $1$		& $1$		& $1$	
	& $1$		& $1$		\\
  $L^2(\vec\sigma_i\cdot\vec\sigma_j)$
			& $-1/3$ 	& $1/9$		& $2/3$	
	& $4/9$		& $-2/9$	\\
  $L^2(\vec\sigma_i\cdot\vec\sigma_j)(\vec\tau_i\cdot\vec\tau_j)$
			& $-1/5$	& $1/25$	& $2/5$	
	& $4/25$	& $-2/25$	\\  
  $(\vec L\cdot\vec S)^2$ 
			& $1/3\hdagb$	& $1/9$		& $2/3\hdagb$
	& $4/9\hdagb$	& $-2/9\hdagb$	\\
  $(\vec L\cdot\vec S)^2(\vec\tau_i\cdot\vec\tau_j)$
			& $1\hdagb$ 	& $1\hdagb$	& $1\hdagb$
	& $1\hdagb$ 	& $1\hdagb$ 	\\
  \hline
  $Q_{12}$			
			& $-1/3$	& $1/9$ 	& $2/3$	
	& $4/9$		& $-2/9$	\\
  $Q_{12}(\vec\tau_i\cdot\vec\tau_j)$
			& $-1/5$	& $1/25$	& $2/5$	
	& $4/25$	& $-2/25$	\\
  \hline\hline
  \end{tabular}
  \end{center}
  \caption{\label{tab:scl_xi} Quark model
scaling factors for the interaction operators for  
two-baryon states formed with $\Xi_c^{'}$, $\Xi_{(cc)}$ hyperons
and nucleons ($N$).
The superscript $\hdagb$ indicates lack of an unambiguous 
scaling factor. For these cases the largest scaling factor
was chosen.}
\vspace{10pt}
\end{table}

The total spin operator $\vec S=(\vec \sigma_i+\vec \sigma_j)/2$ does 
not yield an unambiguous scaling factor,
when the matrix elements contain terms with different scaling behavior.
In the first and the third column of Table~\ref{tab:scl_xi}
(bound states 
with nucleons), the scaling factors for all operators, which
contain the total spin-operator $\vec S$, have 
been approximated by the largest scaling factor in the
corresponding
expression. In the other columns (bound states of two hyperons) 
only the scaling factors containing $(\vec L\cdot \vec S)^2$ had to be
approximated in the same way. 

The scaling factors in Table~\ref{tab:scl_xi} reveal that the
interaction between $\Xi_c^{'}$ and $\Xi_{cc}$ hyperons 
differs qualitatively from the nucleon-nucleon interaction
in that the spin-independent central
interaction, which contains most of the intermediate
range attraction, and the strong short range repulsion
is weaker by an order of magnitude. This weakening of
the short range repulsion increases the relative
importance of the other short range interaction 
components, which are of little significance for --- and
as a consequence not very well constrained by --- 
low energy nucleon-nucleon 
scattering observables. As these other short
range components vary significantly between the
different nucleon-nucleon interaction models, this
will cause a considerable theoretical uncertainty in
the binding energies of the two-baryon systems
when calculated with different interaction models.

\begin{figure}
\begin{center}
\includegraphics[scale=0.5]{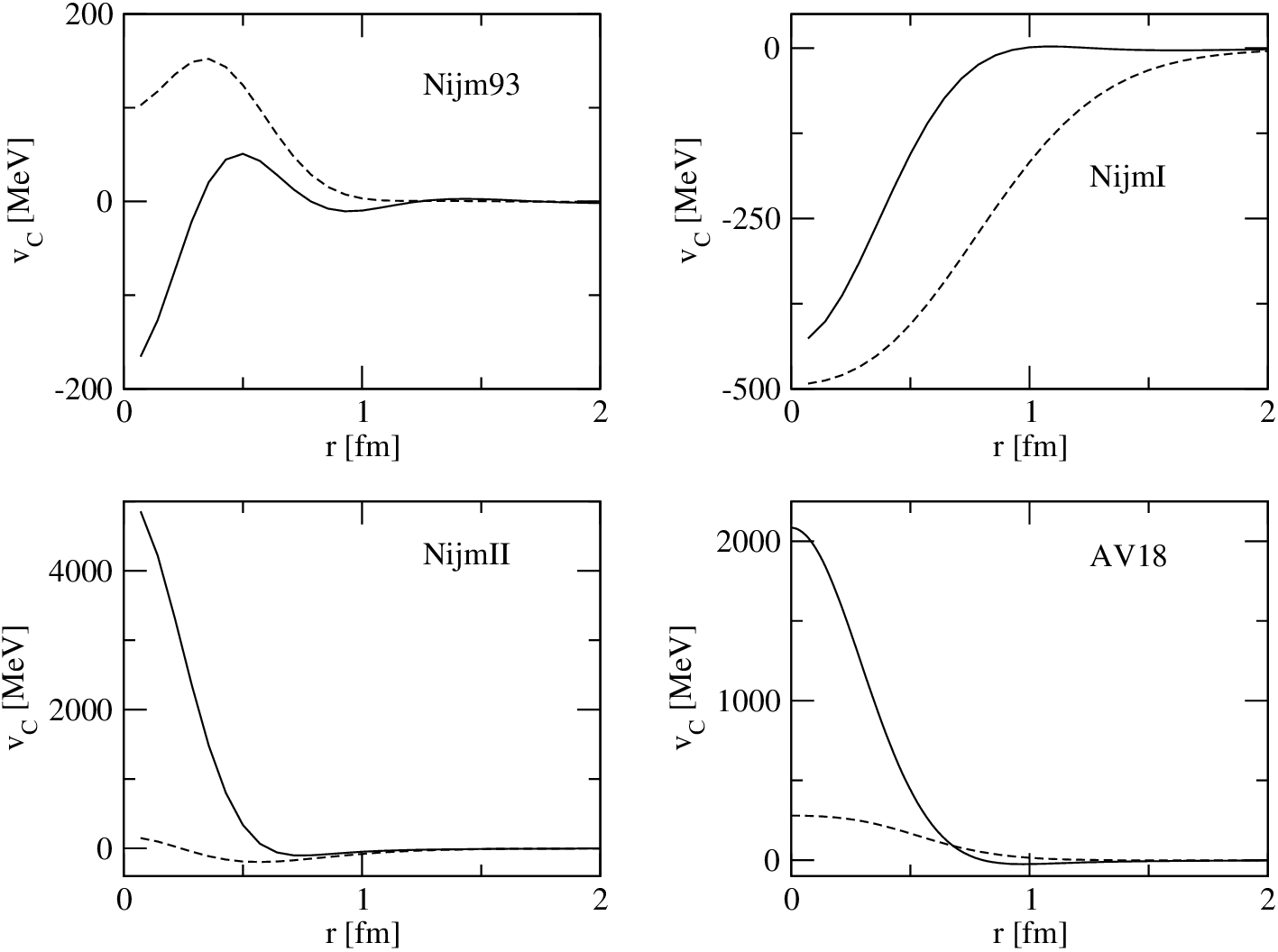}
\end{center}
\caption{\label{fig:vc} Central components of the potentials 
($v_c=v_1\cdot 1 +v_{\tau\tau}\vec\tau_i\cdot\vec\tau_j$) 
for isospin 0. Solid and dashed lines show the deuteron 
potentials and 
the potentials rescaled with the quark model scaling factors for 
the $\Xi_{cc}-N$ system, respectively.}
\end{figure}

The weakening of the short range repulsion is illustrated
for the two non-local Nijmegen potentials Nijm93 and NijmI,
and for the two local potentials NijmII and AV18
in Fig.~\ref{fig:vc}.
In the figure the matrix elements for the isospin
0 state of the central interaction potential
component $v_1(r)-3 v_{\tau\tau}(r)$ is plotted both
for the nucleon-nucleon and the $\Xi_{cc}-N$ systems.
In the latter the quark model rescalings in 
Table~\ref{tab:scl_xi} have been taken into account.
In the case of the Nijm93 interaction model the
short range attraction in the nucleon-nucleon
system is replaced by short range repulsion in the
$\Xi_{cc}-N$ system. In the case of the NijmI
interaction model this potential matrix element
becomes entirely attractive in the case of the
$\Xi_{cc}-N$ system, while in the case of the 
two (local) potentials NijmII and AV18 the short range
repulsion is almost wiped out. 

The two-baryon states of $\Xi_c^{'}$ and $\Xi_{cc}$ baryons 
differ from the deuteron in that the long range pion
exchange interaction is also weaker by more than
an order of magnitude. Due to this weakening
this interaction is not the main source of binding, as
it is in the case of the deuteron.

In Table~\ref{tab:xin_nomc} the binding energies for
isospin 0 combinations of $\Xi_c^{'}-N$ and $\Xi_{cc}-N$
states --- calculated by solving the Schr\"odinger 
equation with the rescaled versions of the three 
Nijmegen~\cite{nijm} and AV18~\cite{AV18}
models for the nucleon-nucleon interaction 
(without the 
isospin breaking electromagnetic components, which
depend on the baryon charge) --- are
listed. For comparison, the corresponding binding energies
that are obtained with the rescaled class of AVn'
interactions are also listed in the table. To obtain these
results only the baryon mass independent scaling
factors from Table~\ref{tab:scl_xi} were used.

The very large binding energies obtained with the
rescaled AV18 interaction models are notable. These
arise from the strong $\vec L^2$ interaction component
in the AV18 potential.  
 
Due to the strongly attractive central component of the 
rescaled NijmI interaction, unrealistic deeply bound
states with binding energies
in the range of $1-10\,$GeV are found with that interaction
model. For some of the
other systems, that will be discussed later, the rescaled
Nijm93 interaction shows a similar behavior. 
In the tables such 
results will be denoted by the entry $***$. 

\begin{table}
\begin{center}
\begin{tabular}{c|cc}
\hline\hline
	& $\Xi_c^{'}-N$	& $\Xi_{cc}-N$	\\
\hline
\small{Potential} 
	& \multicolumn{2}{c}{\small{Binding Energy [MeV]}}\\
\hline\hline
Nijm93	& ---		& ---		\\
NijmI  & *** 		& ***		\\
NijmII & $-13.8$	& $-34.5$	\\
\hline
AV18	& $-17.2$	& $-387.0$	\\
AV8'	& ---		& $-1.0$	\\
AV6'	& ---		& ---		\\
AV4'	& $-3.0$	& ---		\\	
AVX'	& $-0.8$	& $-0.01$	\\
AV2'	& $-0.3$	& $-0.6$	\\
AV1'	& ---		& ---		\\
\hline\hline
\end{tabular}
\end{center}
\caption{\label{tab:xin_nomc} Binding energies of 
$\Xi_c^{'}-N$-type states with 
  $I_\mathrm{tot}=0$ calculated with different
interaction models and the
quark model scaling factors in Table~\ref{tab:scl_xi}.
The entry *** indicates unrealistic deeply
bound states.}
\end{table}

In Table~\ref{tab:xixi_nomc} the corresponding calculated
binding energies for two-baryon states with isospin 0
of the form $\Xi-\Xi_{cc}$ and $\Xi_c^{'}-\Xi_{cc}$ as well
as $\Xi_{cc}-\Xi_{bb}$ are listed. Here the mass of the
$\Xi_{bb}$ was taken to be 9 GeV.
The rescaled AV18 potential implies substantial binding energies
also in the case of these states. 
Concerning the results obtained with the Nijmegen potentials, 
it can be seen that the two non-local versions, Nijm93 and NijmI, 
do not predict bound states and in most cases produce unrealistic 
results. The local version, NijmII, gives rise to bound states.
These binding energies are considerably smaller than
those obtained with the AV18 interaction.

\begin{table}
\begin{center}
\begin{tabular}{c|cccccc}
\hline\hline
	&$\Xi-\Xi_{cc}$	& $\Xi_c^{'}-\Xi_c^{'}$	
					& $\Xi_c^{'}-\Xi_{cc}$
							&
 $\Xi_{cc}-\Xi_{cc}$	& $\Xi_{cc}-\Xi_{bb}$	& $\Xi_{bb}-\Xi_{bb}$	\\
\hline
\small{Potential} &		 \multicolumn{6}{c
}{\small{Binding Energy [MeV]}} 					
	\\
\hline\hline
Nijm93 & ---		& ---		& ---		& ***	
		& ***			& ***			\\
NijmI  & ***		& ***		& ***		& ***	
		& ***			& ***			
	\\
NijmII & $-56.2$	& $-71.0$	&$-87.1\hdag$ 	& $-86.4$
		& $-102.3\hdag$		& $-123.2\hdag$		\\
\hline
AV18	& $-174.0$	& $-457.0$	& $-757.2\hdag$	& $-456.4\hdag$	
	& $-601.0\hdag$		& $-780.1\hdag$		\\
AV8'	& ---		& ---		& ---		& ---		
	& $-0.1$		& $-3.0$		\\
AV6'	& ---		& $-0.7$	& $-0.01$	& $-1.0$	
	& $-5.2$		& $-14.6$		\\
AV4'	& $-12.0$	& $-24.5$	& $-20.9$	& $-29.8$	
	& $-41.3$		& $-57.6\hdag$		\\
AVX'	& $-4.5$	& $-9.5$	& $-11.3$	& $-14.6$	
	& $-21.6$		& $-32.6\hdag$ 		\\
AV2'	& $-4.6$	& $-12.8$	& $-18.1$	& $-25.8$	
	& $-47.7\hdag$		& $-87.8\hdag$		\\
AV1'	& ---		& ---		& ---		& ---		
	& ---			& $-0.1$		\\
\hline\hline
\end{tabular}
\end{center}
\caption{\label{tab:xixi_nomc} Binding energies of
two baryon states of $\Xi$, $\Xi_c^{'}$ and
$\Xi_{cc}$ hyperons with
$I_\mathrm{tot}=0$ as obtained with the quark model
scaling factors in Table~\ref{tab:scl_xi}.
The superscript
$\hdag$ indicates that more than one bound state exists.}

\end{table}

\begin{figure}
\begin{center}
\includegraphics[scale=0.5]{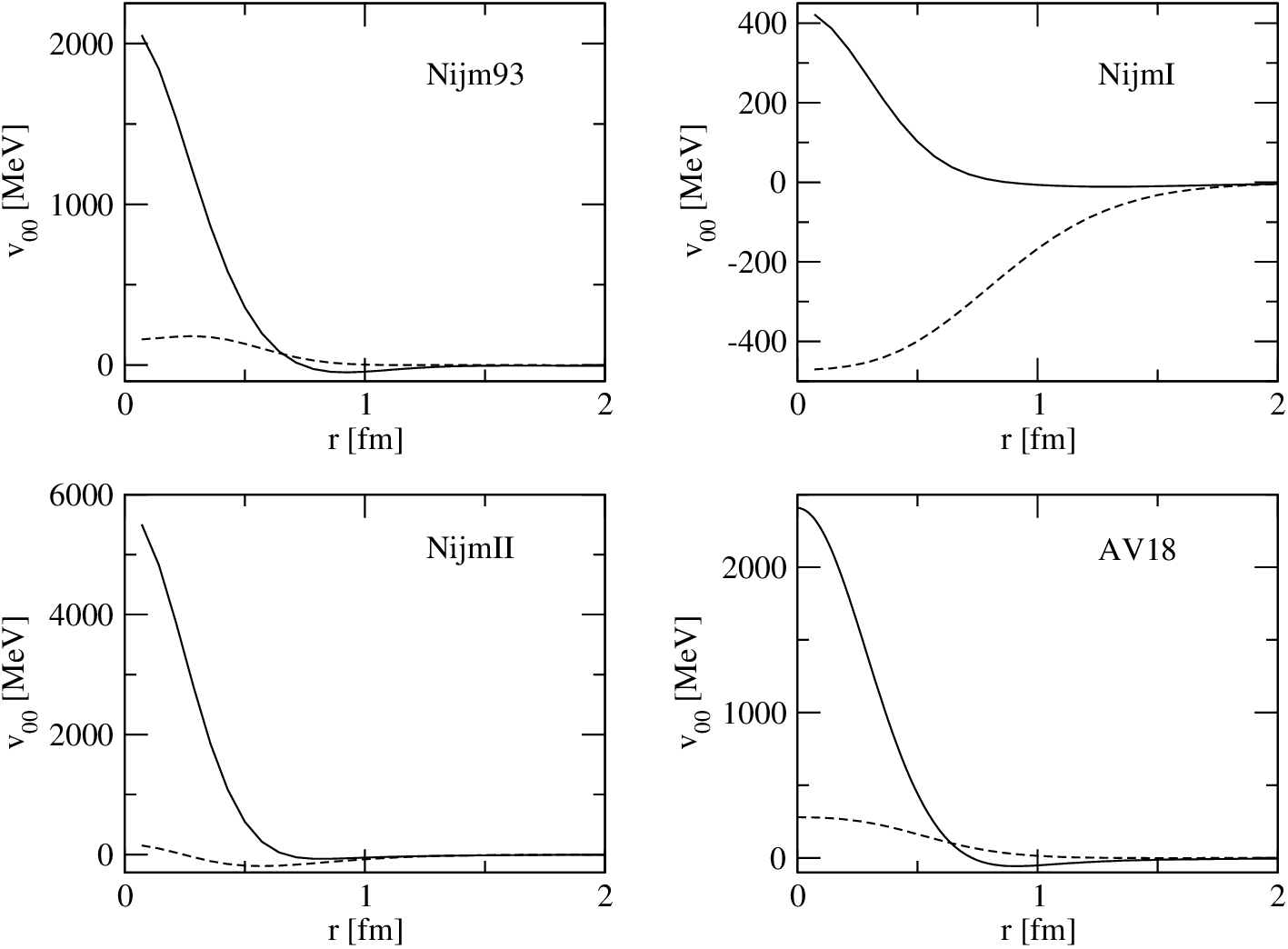}
\end{center}
\caption{\label{fig:v00}Full potentials in the $S-$wave 
channel ($v_{00}$). Solid and dashed lines correspond to 
the deuteron potentials
and the rescaled potentials for the
$\Xi_{cc}-N$ system in the isospin 0 state, respectively.}
\end{figure}

\begin{figure}
\begin{center}
\includegraphics[scale=0.5]{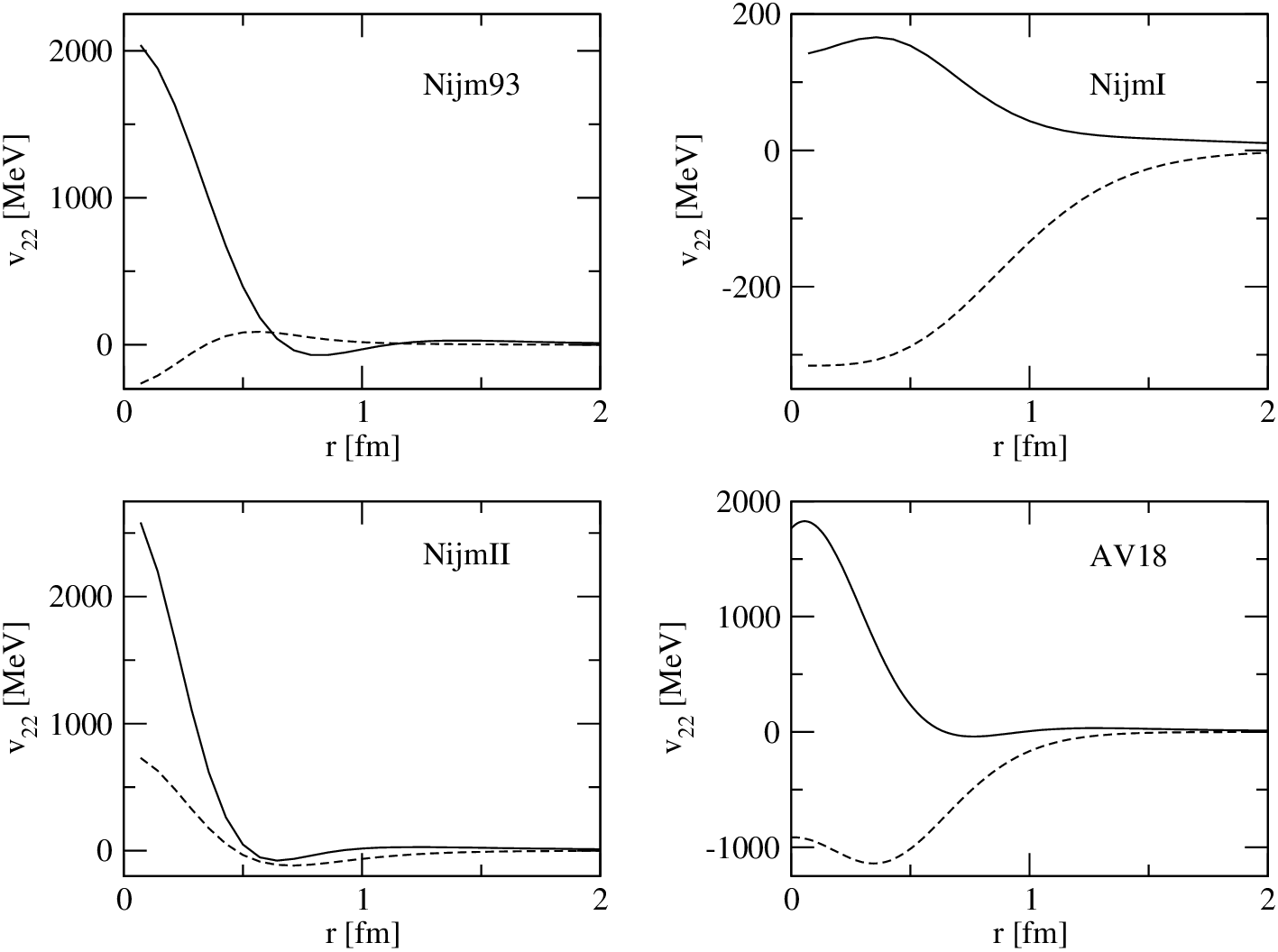}
\end{center}
\caption{\label{fig:v22}Full potentials in the 
$D-$wave channel 
($v_{22}$). Solid and dashed lines depict the deuteron potentials 
and the rescaled potentials for the
$\Xi_{cc}-N$ system with isospin 0, respectively.}
\end{figure}

The substantial spread in the calculated binding energies 
is a direct consequence of the large differences in the
short range parts of the different potential models,
which is accentuated in the case of the $S-$states.
In Figs.~\ref{fig:v00} and \ref{fig:v22} the diagonal
$S-$ and $D-$state potentials for the $\Xi_{cc}-N$
systems are shown for the different potential models.
The NijmI potential model has a very attractive
$D-$state interaction, which leads to the unrealistically
large calculated binding energies. A comparison to the
results that are obtained with the earlier Paris
potential model \cite{lacombe} emphasizes this point.
This potential model leads to considerably smaller
values for the binding energies: no binding for the
the $\Xi_{cc}-\Xi_{cc}$ and the
$\Xi_{cc}-\Xi_{bb}$ systems and only  2 MeV for the
$\Xi_{bb}-\Xi_{bb}$ system. That the substantial
cancellation between the strongly attractive and
repulsive components in the NijmI potential model
may lead to peculiar results in the extension to 
hyperons has been noted before \cite{rijken}.

The large binding energies that are associated
with the $\vec L^2$ component of  
the rescaled AV18 potential suggest
that the quark model scaling factors in Table~\ref{tab:scl_xi} 
may be oversimplified in that they do not take into
account the possibility of baryon mass dependence
in the scaling factors of the interaction components, which
depend explicitly on angular momentum. 
The scaling factors in Table~\ref{tab:scl_xi} are obtained
algebraically by application of the quark model
wave functions. They therefore do not take any account
of the fact that in dynamical models of the nucleon-nucleon
interaction, the components which contain the angular
momentum operator $\vec L$, always contain an inverse
power of the nucleon mass. This dependence on an inverse
power of the nucleon mass plays a crucial role in the
determination of the scaling of the nuclear 
interaction in the large $N_C$ limit~\cite{kaplan,riska}. 

In a quark model based
derivation of the nucleon-nucleon interaction, it will,
however, be the constituent quark and not the baryon masses
which arise in the angular momentum dependent components.
As in the present calculation, where only the interactions
between the light flavor quarks are considered, the
scaling factors in the table do not depend on the baryon
mass. In order to investigate the numerical sensitivity
to this issue, the possibility of a baryon mass dependence
for the scaling factors, which are associated with the
angular momentum dependent operators in (\ref{opv}), was 
investigated.
  
For the $\vec L\cdot\vec S$, $\vec L\,^2$ and  
$(\vec L\cdot\vec S)^2$ potential components for baryons
of unequal mass we consider
the additional mass dependent scaling factor:
    \beq
      1 \quad\rightarrow\quad 
      \frac{1}{2}\left[ 1+\left(\frac{m}{M}\right)^2  \right]\, ,
\label{massdep1}
    \eeq
and for interactions that involve the quadratic spin-orbit
interaction operator $Q_{12}$ the mass dependent scaling
factor is taken to be:
    \beq
      1 \quad\rightarrow\quad \left(\frac{m}{M}\right)^2\, .
\label{massdep2}
    \eeq

Here $m$ is the mass of the light baryon and $M$ is the mass 
of the heavy baryon. The scaling factor \ref{massdep1}
is that for a scalar interaction, and is employed here only
for a qualitative estimate.

In order to study the significance of this issue, the
binding energies of the $N-\Xi_c^{'}$ and
$N-\Xi_{cc}$ systems, as obtained when the scaling
factors for those components are allowed to depend
on mass as in (\ref{massdep1}) and (\ref{massdep2}),
are shown in Table~\ref{tab:xin_mc}.
With mass dependent scaling the AV18 potential does not
lead to any bound state. In contrast the NijmII interaction
model result for the binding energies is
only slightly smaller when the mass dependence of the
scaling factors is taken into account. This is a
consequence of the fact that the angular momentum
dependence of the NijmII interaction model is weak
in comparison to that of the AV18 interaction. 

\begin{table}
\begin{center}
\begin{tabular}{c|cc}
\hline\hline
	& $\Xi_c^{'}-N$	& $\Xi_{cc}-N$	\\
\hline
\small{Potential} 
	& \multicolumn{2}{c}{\small{Binding Energy [MeV]}} \\
\hline\hline
Nijm93	& ---		& ---		\\
NijmI  & *** 		& *** 		\\
NijmII & $-11.2$	& $-19.4$	\\
\hline
AV18	& ---		& ---		\\
AV8'	& ---		& ---		\\
AV6'	& ---		& ---		\\
AV4'	& $-3.0$	& ---		\\
AVX'	& $-0.8$	& $-0.01$	\\
AV2'	& $-0.3$	& $-0.6$	\\
AV1'	& ---		& ---		\\
\hline\hline
\end{tabular}
\end{center}
\caption{\label{tab:xin_mc} Binding energies of 
$\Xi_c^{'}-N$ and $\Xi_{cc}-N$ with 
  $I_\mathrm{tot}=0$ as obtained 
  with the quark model scalings in Table~\ref{tab:scl_xi}
 with the additional mass 
factors in Eqs.~(\ref{massdep1}) and (\ref{massdep2})
in the angular momentum dependent terms.}
\end{table}

In the case of the isospin 0 two-baryon states of the
form $\Xi-\Xi_{cc}$, $\Xi_c^{'}-\Xi_{cc}$ and $\Xi_{cc}-\Xi_{bb}$
the results are similar, as shown in Table~\ref{tab:xixi_mc}.
The binding energies calculated with
the AV18 are considerably smaller, when the mass 
dependence of the scaling factors is taken into 
account. This result indicates that more confidence should
be given to the results that are obtained with the NijmII
potential model, which has a weaker dependence on angular
momentum than the AV18 interaction. This conclusion is
also supported by the fact that the angular momentum
independent AVn' potentials give rise to far smaller
binding energies than the AV18 interaction.
The results for the two non-local Nijmegen potentials 
Nijm93 and NijmI remain unrealistic even when the
mass dependent scaling is considered. This seems to indicate
that these potentials  are not very well suited for the 
rescaling procedure used here.

These numerical results suggest that deuteron-like
weakly bound states are formed by nucleons and both
$\Xi_c^{'}$ and $\Xi_{cc}$ hyperons. Somewhat
more strongly bound deuteron-like states are
likely to be formed by $\Xi_c^{'}$ and
$\Xi_{cc}$ hyperons and such bound states are
also likely to be formed by $\Xi$ hyperons and the
charm $\Xi_c^{'}$ and double-charm $\Xi_{cc}$ hyperons. 
The corresponding states formed with the beauty hyperons 
$\Xi_{b}^{'}$ and (yet to be discovered) $\Xi_{bb}$ will
be deeply bound. As noted above, the NijmII interaction
should be expected to give the most realistic estimates
for the binding energies.

\begin{table}
\begin{center}
\begin{tabular}{c|ccc}
\hline\hline
			& $\Xi-\Xi_{cc}$	& $\Xi_c^{'}-\Xi_{cc}$
	& $\Xi_{cc}-\Xi_{bb}$	\\
\hline
\small{Potential} 	& 	\multicolumn{3}{c}{\small{Binding Energy
 [MeV]}} 		\\
\hline\hline
Nijm93 		& ---			& ---		
	& ***			\\
NijmI  		& ***			& ***		
	& ***			\\
NijmII			& $-56.2$		& $-86.2\hdag$
	& $-102.3\hdag$	\\
\hline
AV18			& ---			& $-438.0$	
	& $-159.0$		\\
AV8'			& ---			& ---		
	& $-0.1$		\\
AV6'			& ---			& $-0.01$	
	& $-5.2$		\\
AV4'			& $-12.0$		& $-20.9$	
	& $-41.3$		\\
AVX'			& $-4.5$		& $-11.3$	
	& $-21.6$		\\
AV2'			& $-4.6$		& $-18.1$	
	& $-47.7\hdag$	\\
AV1'			& ---			& ---		
	& ---			\\
\hline\hline
\end{tabular}
\end{center}
\caption{\label{tab:xixi_mc} The calculated binding energies
of $\Xi-\Xi_{cc}$, $\Xi_c^{'}-\Xi_{cc}$ and
$\Xi_{cc}-\Xi_{bb}$ states with 
  $I_\mathrm{tot}=0$
a obtained with the quark model scaling
factors in Table~\ref{tab:scl_xi} with the additional mass 
factors in Eqs.~(\ref{massdep1}) and (\ref{massdep2})
in the angular momentum dependent terms.}
\end{table}

Above, only the diagonal interactions in the two-baryon states
formed by heavy hyperons have been included. Transitions
between different combinations of charm hyperons are
however possible through charm exchange interactions. Because
of their short range they will be much weaker than the
diagonal interactions that are mediated by light flavor
exchange mechanisms.

Two-baryon states formed by $\Xi_{cc}$ may couple to
$\Lambda_c-\Omega_{ccc}$ states. If the total mass
of the latter is lower than the mass of the bound
$\Xi_{cc}-\Xi_{cc}$ state, the latter state will only
be metastable. While the mass of the $\Omega_{ccc}$
is unknown experimentally, several quark models 
suggest that the mass of the $\Lambda_c-\Omega_{ccc}$ 
state may be lower by some 130-160 MeV \cite{rich,silv}.
In this case the AV18 interaction, without mass dependent 
scalings, leads to 
$\Xi_{cc}-\Xi_{cc}$ bound states but the NijmII interaction
only to a metastable state.

In the case of the isospin 0 state of the
$\Xi_{cc}-N$ system the charm exchange
interactions couple it to the $\Lambda_c^+ -\Lambda_c^+$
system, which has a
mass that is $\sim$ 200 MeV higher than that of the
$\Xi_{cc}-N$ bound states. Consequently the predicted
$\Xi_{cc}-N$ bound states are bound rather than metastable.

The isospin 0 state of the $\Xi^{'}_{c}-N$ system
is also bound. It is coupled to the $\Sigma-\Sigma_c$
system. The states of this system have $\sim$ 130 MeV
higher masses than the $\Xi^{'}_{c}-N$ states.

In contrast the calculated ``bound'' states of two
$\Xi_c^{'}$ hyperons, which may couple to the $\Xi-\Xi_{cc}$
state by charm exchange will be only metastable, as the
mass of the latter is lower by $\sim 370$ MeV. 
The states of the latter
are consequently bound and not metastable.

\section{Two-baryon states with isospin 1 charm hyperons}
\label{sec:isoone}

\subsection{Baryon-baryon
interactions with isospin 1 baryons}

The scaling approach that is presented here can be readily extended 
to isospin 1 baryons. As long as only baryons with spin 1/2 are 
considered it is only the isospin dependent part of the model that
has to be modified. 

The quark model scaling factors can be calculated as in the previous
section. The scaling factors for the isospin dependent interaction
components refer to the scaling relative to the sum over the isospins
of the constituent quarks, $\sum_{q=1}^3 \vec\tau^q$ and not relative
to the baryon isospin operators $\vec\tau$ or $\vec T$. Therefore,
the procedure applies independently of the isospin of the considered
baryons. 

The isospin 1 baryon operators $T^{1,2,3}$ satisfy the
$SU(2)$ algebra $[T^p,T^q]=i\epsilon_{pqr}T^r$, without any factor
$\frac{1}{2}$ on the right hand side as in the case of
the Pauli isospin matrices
($[\frac{1}{2}\tau^p,\frac{1}{2}\tau^q]=i\epsilon_{pqr}\frac{1}{2}\tau^r$).
Hence, the isospin operator $\vec T$ for baryons with isospin 1 
corresponds to $2\vec T=\sum_q {\vec\tau\,}^q$. The factor of 2 on
the l.h.s. is included in the scaling
factors employed here. This is important when the NN potentials are
be expressed in terms of baryon operators. The isospin $\frac{1}{2}$
operators $\vec\tau_i\cdot\vec\tau_j$ must then be 
replaced by $\vec T_i\cdot\vec\tau_j$ or $\vec T_i\cdot\vec T_j$,
respectively. The ``missing'' factors $2$ or $4$ are generated
by the scaling factors.

The scalings of the matrix elements for the light quark operators
between $\Sigma_c^{++}$ or $\Sigma_c^{0}$ hyperons and nucleons in the
quark model are:
\bea
	&&\matelsym{\signpp}{\sum_q 1^q}      
		= \frac{2}{3}\matelsym{N}{\sum_q 1^q} \, ,\nnl
	&&\matelsym{\signpp}{\sum_q\sigma_i^q}  
		= \frac{4}{3}\matelsym{N}{\sum_q\sigma_i^q}
\, , \nnl
	&&\matelsym{\signpp}{\sum_q\tau_i^q}    
		= 2 \matelsym{N}{\sum_q\tau_i^q} \, ,  \nnl
	&&\matelsym{\signpp}{\sum_q\sigma_i^q\tau_j^q} 
		= \frac{4}{5}\matelsym{N}{\sum_q\sigma_i^q\tau_j^q}
\, . \label{eq:sclppn}
\eea
The isospin-z projection of the $\sigp$ hyperon is zero. Thus, the
isospin dependent scalings are different for this charge state:
\bea
	&&\matelsym{\sigp}{\sum_q 1_2^q}      
		= \frac{2}{3}\matelsym{N}{\sum_q 1_2^q}\, , \nnl
	&&\matelsym{\sigp}{\sum_q\sigma_i^q}  
		= \frac{4}{3}\matelsym{N}{\sum_q\sigma_i^q}\, , \nnl
	&&\matelsym{\sigp}{\sum_q\tau_i^q}    
		= 0  \cdot  \matelsym{N}{\sum_q\tau_i^q} \, ,  \nnl
	&&\matelsym{\sigp}{\sum_q\sigma_i^q\tau_j^q} 
		= 0 \cdot \matelsym{N}{\sum_q\sigma_i^q\tau_j^q}\, .
		\label{eq:sclp}
\eea
Binding energies will be calculated for isospin eigenstates that 
are constructed from the charge states. Therefore, non-diagonal 
matrix elements also have to be considered. The scalings for the 
matrix elements between $\signpp$ and $\sigp$ hyperons and matrix
elements between proton and neutron states are:
\bea
	&&\matel{\signpp}{\sum_q \tau_i^q}{\sigp}
		= \pm\sqrt{2} \matel{p}{\sum_q \tau_i^q}{n}\, , \nnl
	&&\matel{\signpp}{\sum_q \sigma_i^q\tau_j^q}{\sigp}
		= \pm\frac{2\sqrt{2}}{5}\matel{p}{\sum_q 
\sigma_i^q\tau_j^q}{n} \, . \label{eq:sclnd}
\eea

>From the differences in the isospin dependent scalings in 
Eqs.~(\ref{eq:sclppn})-(\ref{eq:sclnd}) it follows that the full
 scaling behavior of isospin eigenstates cannot be described by
 the quark model scaling factors alone. The crucial point is
 that the structure of the isospin eigenstates changes when 
isospin 1/2 baryons are replaced by isospin $1$ baryons.
 There are then three ($\sigpp$, $\sigp$, $\sign$) 
charge states instead of
 two ($p$, $n$). This change of the 
structure --- which manifests in different Clebsch-Gordan
 coefficients --- must be seen as a part of the scaling
 procedure. The quark model scaling factors are derived 
for the individual charge states.  They have no information
 about the arrangement of the charge states into isospin 
eigenstates.  The total scaling will thus be a combination 
of the quark model scaling factors and the Clebsch-Gordan 
coefficients, which has to be determined individually 
for each isospin eigenstate.

In the following sections it is shown that the full scaling of a 
isospin dependent interaction component can be split up into a
quark model scaling factor and another scaling factor that is 
related to the explicit value of the matrix element of the 
two-baryon state considered. The matrix element relations
for the relevant isospin eigenstates are:
\bea
	\matelsym{1;\pm m}{\vec\tau_i\cdot\vec\tau_j} 
&=&1\, ,\quad m=0,1 \, , \nonumber \\ 
	\matelsym{0;0}{\vec\tau_i\cdot\vec\tau_j}&=&-3 \, ,
 \label{eq:matelnn}
\eea
\bea
	\matelsym{\frac{3}{2};m}{\vec T_i\cdot\vec \tau_j}&=&1 
 \, ,
\quad m=\frac{1}{2},\frac{3}{2} \, ,\nonumber \\
	\matelsym{\frac{1}{2};\pm\frac{1}{2}}{\vec T_i\cdot\vec
 \tau_j} &=& -2
\, , \label{eq:matelsign}
\eea
\bea
	\matelsym{2;\pm m}{\vec T_i \cdot \vec T_j} 
&=&  1\, , \quad m=0,1,2 \, ,\nonumber  \\
	\matelsym{1;\pm m}{\vec T_i \cdot \vec T_j} 
&=& -1\,  , \quad m=0,1 \, ,\nonumber  \\
	\matelsym{0;0}{\vec T_i \cdot \vec T_j} &=& -2 \, .
 \label{eq:matelsigsig}
\eea

Note that the composition of the eigenstates could be ignored when
only isospin 1/2 baryons were considered. The isospin eigenstates
of nucleons and $\Xi_{(cc,bb)}$ baryons are constructed identically,
i.e., using the same Clebsch-Gordan coefficients. Therefore, the
full scaling information is contained in the quark model
scaling factors.

\subsection{Bound states of two $\Sigma_c$ hyperons}

The $\Sigma_c^{++}$ and the $\Sigma_c^{0}$ 
hyperons form isospin symmetric two-body states with 
isospin 2 and one antisymmetric state with isospin 1. 
These are explicitly:
\bea
    &&\wf{2;2}=\wf{\sigpp\sigpp}\, , \quad
      \wf{2,-2}=\wf{\sign\sign}\, , \nonumber \\
    &&
      \wf{1,0}=
        \frac{1}{\sqrt{2}}\left[\wf{\sigpp\sign}
	-\wf{\sign\sigpp}\right]\, . \label{eq:wf10}
\eea
The states $\wf{2;0}$, $\wf{2;\pm1}$, $\wf{0;0}$ and 
$\wf{1;\pm 1}$ are linear
combinations of
all the $\Sigma_c$ charge states. Their explicit expressions
are:
\bea
	&&\wf{2;0} = \frac{1}{\sqrt{6}}\left[\wf{\sigpp\sign}
		+2\wf{\sigp\sigp}+\wf{\sign\sigpp}\right]\, , \\
	&&	\wf{2;1}=\frac{1}{\sqrt{2}}\left[
		\wf{\sigpp\sigp}+\wf{\sigp\sigpp}
		\right]\, , \\
	&&	\wf{2;-1}=\frac{1}{\sqrt{2}}\left[
			\wf{\sign\sigp}+\wf{\sigp\sign}
		\right]\, , 			
\eea
and
\bea
&&  	\wf{1; 1}=\frac{1}{\sqrt{2}}\left[
			\wf{\sigpp\sigp}-\wf{\sigp\sigpp}
		\right]\, , \\
&&  	\wf{1;-1}=\frac{1}{\sqrt{2}}\left[
			\wf{\sigp\sign}-\wf{\sign\sigp}
		\right]\, , \\
			&&\wf{0;0} = \frac{1}{\sqrt{3}}\left[
			\wf{\sigpp\sign}-\wf{\sigp\sigp}
			+\wf{\sign\sigpp}\right]\, , \label{eq:wf00}
\eea
respectively The isospin independent scaling factors for those states,
which can be derived directly from Eqs.~(\ref{eq:sclppn}), are listed in
 Table~\ref{tab:scl_sigppn}. The full scaling behavior of the isospin 
dependent interaction components is found by calculating the scaling
 between each of the two-baryon state matrix elements and the corresponding
 isospin symmetric or antisymmetric two-nucleon matrix element.

The simplest case is the $\wf{2;2}$ state, since the structure of this 
isospin eigenstate is identical to the structure of the $\wf{1;1}$ state
 that is formed of two isospin 1/2 baryons. Using Eqs.~(\ref{eq:sclppn})
 it is found:
\bea
	\matelsym{2;2}{\sum_p\vec\tau^p_i\cdot\sum_q\vec\tau^q_j}&=
&\matelsym{\sigpp}{\sum_q\vec\tau^q}\cdot\matelsym{\sigpp}{\sum_q\vec\tau^q}
 \nnl
	&=&2	\matelsym{p}{\sum_q\vec\tau^q}\cdot 2 \matelsym{p}
{\sum_q\vec\tau^q} \nnl
	&=& 4 \matelsym{1;1}{\sum_p\vec\tau^p_i\cdot\sum_q
\vec\tau^q_j}_{NN} \label{eq:scl22}
\eea
The state $\wf{0;0}$ mixes the charge states. Thus,
 all the scalings from Eqs.~(\ref{eq:sclppn})-(\ref{eq:sclnd}) enter
 into the two-body matrix elements:
\bea
	\matelsym{0;0}{\sum_p\vec\tau^p_i\cdot\sum_q\vec\tau^q_j}
	&=&\frac{2}{3}\left[
		\matelsym{\sigpp}{\sum_q\vec\tau^q}
\cdot\matelsym{\sign}{\sum_q\vec\tau^q} \right. \nnl
		&&-\matel{\sigpp}{\sum_q\vec\tau^q}{\sigp}
\cdot\matel{\sign}{\sum_q\vec\tau^q}{\sigp} \nnl
		 &&\left. -\matel{\sigp}{\sum_q\vec\tau^q
}{\sigpp}\cdot\matel{\sigp}{\sum_q\vec\tau^q}{\sign} 
	\right] \nnl
	&=&\frac{2}{3}\left[2\matelsym{p}{\sum_q\vec\tau^q}
\cdot 2\matelsym{n}{\sum_q\vec\tau^q} \right. \nnl
	&&\left. -2\sqrt{2}\matel{p}{\sum_q\vec\tau^q}{n}
\cdot\sqrt{2}\matel{n}{\sum_q\vec\tau^q}{p} \right] \nnl
	&=&4\cdot\frac{2}{3}\matelsym{0;0}{\sum_p\vec\tau^p_i
\cdot\sum_q\vec\tau^q_j}_{NN}\, . \label{eq:scl00}
\eea

This calculation can be repeated analogously for all the states
 (\ref{eq:wf10})-(\ref{eq:wf00}). It is found that the scaling
only depends on the total isospin of each state and not
 on the isospin-z projection. The explicit results for the 
scaling factors are:
\bea
	&&\matelsym{2;\pm m}{\sum_p\vec\tau^p_i\cdot\sum_q\vec\tau^q_j
}=4\cdot 1\matelsym{1;1}{\sum_p\vec\tau^p_i
\cdot\sum_q\vec\tau^q_j}_{NN}\, , \nnl
	&&\matelsym{1;\pm m}{\sum_p\vec\tau^p_i
\cdot\sum_q\vec\tau^q_j}=4\cdot\frac{1}{3}\matelsym{0;0}
{\sum_p\vec\tau^p_i\cdot\sum_q\vec\tau^q_j}_{NN} \, , \nnl
	&&\matelsym{0;0}{\sum_p\vec\tau^p_i
\cdot\sum_q\vec\tau^q_j}=4\cdot\frac{2}{3}\matelsym{0;0}
{\sum_p\vec\tau^p_i\cdot\sum_q\vec\tau^q_j}_{NN}\, . 
\eea

The factor $4$ can be interpreted as the quark model scaling factor
 that is derived from the matrix element scaling in~(\ref{eq:sclppn}).
 The meaning of the factors $1$, $\frac{1}{3}$ and $\frac{2}{3}$
 becomes clear from Eqs.~(\ref{eq:matelnn})-(\ref{eq:matelsigsig}).
 These isospin dependent factors are given by the ratio of the matrix
 elements for the considered two-baryon state and the corresponding 
symmetric or antisymmetric two-nucleon state:
\bea
	&&	1=\frac{\matelsym{2;\pm m}{\vec T_i \cdot \vec T_j} }
{\matelsym{1;1}{\vec\tau_i\cdot\vec\tau_j}_{NN}} \, , \quad m=0,1,2\,
 , \label{eq:isoscl} \\
	&&	\frac{1}{3}=\frac{\matelsym{1;\pm m}{\vec T_i \cdot
 \vec T_j} }{\matelsym{0;0}{\vec\tau_i\cdot\vec\tau_j}_{NN}} \, ,
 \quad m=0,1\, 
	\quad,	\frac{2}{3}=\frac{\matelsym{0;0}{\vec T_i \cdot
 \vec T_j}}{\matelsym{0;0}{\vec\tau_i\cdot\vec\tau_j}_{NN}}\, . \nn
\eea

\begin{table}
  \begin{center}
  \begin{tabular}{c|cc}
  \hline\hline
  			& $\Sigma_c-N$	& $\Sigma_c-\Sigma_c$ 	\\
  \hline
  Operator		& \multicolumn{2}{c}{Scaling factor}	\\
  \hline\hline
  $1$			&	$2/3$	&	$4/9$		\\
  $\vec\tau_i\cdot\vec\tau_j$	
			&	$2$	&	$4$		\\
  $\vec\sigma_i\cdot\vec\sigma_j$ 
			&	$4/3$	&	$16/9$		\\
  $(\vec\sigma_i\cdot\vec\sigma_j)(\vec\tau_i\cdot\vec\tau_j)$
			&	$4/5$	&	$16/25$		\\
  $S_{ij}$		
			&	$4/3$	&	$16/9$		\\
  $S_{ij}(\vec\tau_i\cdot\vec\tau_j)$
			&	$4/5$	&	$16/25$		\\
  $\vec L\cdot\vec S$	
			&$4/3\hdagb$	&	$8/9$		\\
  $\vec L\cdot\vec S (\vec\tau_i\cdot\vec\tau_j)$
			&$2\hdagb$ 	&	$8/5$		\\
  \hline
  $L^2$			
			&	$2/3$	&	$4/9$		\\
  $L^2(\vec\tau_i\cdot\vec\tau_j)$
			&	$2$	&	$4$		\\
  $L^2(\vec\sigma_i\cdot\vec\sigma_j)$
			&	$4/3$	&	$16/9$		\\
  $L^2(\vec\sigma_i\cdot\vec\sigma_j)(\vec\tau_i\cdot\vec\tau_j)$
			&	$4/5$	&	$16/25$		\\
  $(\vec L\cdot\vec S)^2$ 
			& $4/3\hdagb$	&	$16/9\hdagb$	\\
  $(\vec L\cdot\vec S)^2(\vec\tau_i\cdot\vec\tau_j)$
			& $2\hdagb$	& 	$4\hdagb$ 	\\
  \hline
  $Q_{12}$			
			& $4/3$		&	$16/9$ 		\\
  $Q_{12}(\vec\tau_i\cdot\vec\tau_j)$
			& $4/5$		&	$16/25$		\\
  \hline\hline 
  \end{tabular}
  \end{center}
  \caption{\label{tab:scl_sigppn} 
 Quark model
scaling factors for the interaction operators for the  
two-baryon states $\Sigma_c-N$ and $\Sigma_c-\Sigma_c$.
The isospin dependent scaling factors must be multiplied
with the appropriate factor from Eqs.~(\ref{eq:isoscl}) or
(\ref{eq:isoscl2}) to obtain the full scaling factors.
}
\vspace{10pt}
\end{table}

For the other isospin dependent interaction components, like 
$(\vec\sigma_i\cdot\vec\sigma_j)(\vec\tau_i\cdot\vec\tau_j)$,
 the same behavior is found by a similar calculation. The complete scaling
 factors are products of the scaling factors that can be derived from
 Eqs.~(\ref{eq:sclppn}) and the isospin dependent scaling factors
 (\ref{eq:isoscl}). The quark model scaling factors are listed in
 Table~\ref{tab:scl_sigppn}. The scaling factors that could be
 calculated from Eqs.~(\ref{eq:sclp}) and (\ref{eq:sclnd}) do not 
appear explicitly in the total scalings. They have merged with the
 Clebsch-Gordan coefficients to provide an unabiguous scaling
 behavior of the isospin eigenstates.

Note that the $\sigpp-\sigpp$ and $\sign-\sign$ 
light flavor quark interactions
scale from the corresponding
nucleon-nucleon interactions in the same way. The flavor
breaking terms are insignificant as $\sigpp$ 
and $\sign$ have almost the same mass ($\Delta m=0.4$ MeV).
Hence, the results for bound states with
different isospin-z projections will be the same.

The calculated binding energies for the two-charm hyperon
states are listed in Table~\ref{tab:sigsig}. For the symmetric isospin
2 states the Nijm93
interaction and the AV18 interaction lead to bound states,
while the NijmII and most of the AVn' potential models yield no bound
states.
In contrast, the NijmII and many of the AVn' potentials give rise
to bound states for the antisymmetric isospin 1 and 0 states. 
The isospin 0 state is much stronger bound than the isospin 1
states. While the binding energies are rather high  
there is no convergence on the value of the binding energy for
the different potentials.

\begin{table}
\begin{center}
\begin{tabular}{c|ccc}
\hline\hline
	& $\wf{2, m}$
			& $\wf{1,m}$ 	& $\wf{0,0}$ \\
\hline
\small{Potential} 
	& \multicolumn{3}{c}{\small{Binding Energy [MeV]}} 	
					\\
\hline\hline
Nijm93 & $-66.6$	& ---	 	& ***		\\
NijmI  & ---	& ***	 	& ***			\\
NijmII & ---	& $-53.7$	& $-285.8$			\\
\hline
AV18	& $-41.1$	& ---		& ***		\\
AV8'	& ---		& ---		& $-16.1$			\\
AV6'	& ---		& ---		& $-10.8$			\\
AV4'	& ---		& $-7.3$	& $-87.4$		\\
AVX'	& ---		& $-2.8$	& $-53.3$		\\
AV2'	& ---		& $-8.3$	& $-58.5$		\\

AV1'	& $-0.7$	& $-0.7$	& $-0.7$		\\
\hline\hline
\end{tabular}
\end{center}
\caption{\label{tab:sigsig} Calculated binding energy
for $\Sigma_c-\Sigma_c$ states obtained with the
quark model scalings of the interaction operators.}
\end{table}

The bound states of two $\Sigma_c$ 
hyperons found in Table~\ref{tab:sigsig} for some of the
interactions are only metastable. The isospin 2 states
can couple to the lower lying $\Xi_{cc}-\Delta(1232)$ states
that have the same isospin by the short range charm
exchange interaction. In the same way the isospin 1 and 0 
states can couple to the even lower lying $\Xi_{cc}-N$ states.
Even the the strong binding for the isospin 0 states from
the AV18 interaction cannot compensate the large mass
difference of $\sim$ 500 MeV.

In contrast the bound states of $\Sigma_c$ hyperons and 
nucleons that will be discussed in the following section 
are stable. They can only couple to the
$\Lambda_c-\Delta(1232)$ and $\Sigma_c-\Delta(1232)$ 
states which have $\sim$ 120--300 MeV higher masses.

\subsection{$\Sigma_c-N$-type bound states}

The simplest combinations of nucleons and $\Sigma_c$ hyperons
are the two symmetric isospin $3/2$ combinations:
\bea
&&	\wfbig{\frac{3}{2};\frac{3}{2}}
	=\wf{\sigpp p}\, , \quad 
	\wfbig{\frac{3}{2};-\frac{3}{2}}
	=\wf{\sign n}\ .
\eea
The isospin $3/2$ states with $1/2$ and $-1/2$ isospin-z
projections are the two symmetric linear combinations:
\bea
&&	\wfbig{\frac{3}{2};\frac{1}{2}}
                =\frac{1}{\sqrt{3}}\left[\wf{\sigpp n}
		+\sqrt{2}\wf{\sigp p}\right]\, , \\
&&	\wfbig{\frac{3}{2};-\frac{1}{2}}
                =\frac{1}{\sqrt{3}}
		\left[\sqrt{2}\wf{\sigp n}+\wf{\sign p}\right]\, .
\eea
In addition there are the two antisymmetric isospin $1/2$
combinations:
\bea
&&	\wfbig{\frac{1}{2};\frac{1}{2}}
                =\frac{1}{\sqrt{3}}\left[\sqrt{2}\wf{\sigpp n}
		-\wf{\sigp p}\right] \, ,\\
&&	\wfbig{\frac{1}{2};-\frac{1}{2}}
                =\frac{1}{\sqrt{3}}
	        \left[\wf{\sigp n}-\sqrt{2}\wf{\sign p}\right]\, .
\eea

As in the previous section, the isospin independent scaling factors can be
 derived directly from Eqs.~(\ref{eq:sclppn}). The full scaling behavior
 of the isospin dependent interaction components has to be determined by
 calculating the scaling between each of the two-body matrix elements and
 the corresponding isospin symmetric or antisymmetric two-nucleon matrix
 element. Again, the scalings will not depend on the isospin-z projections
 but only on the total isospin of each state. From calculations similar
 to those in Eqs.~(\ref{eq:scl22}) and (\ref{eq:scl00}) full scaling 
factors are found:
\bea
	&&\matelsym{\frac{3}{2};\pm m}{\sum_p\vec\tau^p_i
\cdot\sum_q\vec\tau^q_j}
	=2\cdot  1\matelsym{1;1}{\sum_p\vec\tau^p_i
\cdot\sum_q\vec\tau^q_j}_{NN}\, , \nnl
		&&\matelsym{\frac{1}{2};\pm m}{\sum_p
\vec\tau^p_i\cdot\sum_q\vec\tau^q_j}
	=2\cdot\frac{2}{3}\matelsym{0;0}{\sum_p\vec
\tau^p_i\cdot\sum_q\vec\tau^q_j}_{NN}\, . 
\eea
The factor $2$ on the r.h.s. of these equations 
is the quark model scaling factor derived
 from the matrix element scaling in Eqs.~(\ref{eq:sclppn}). 
 Analogously to Eqs.~(\ref{eq:isoscl}) the isospin dependent
 factors 1 and $\frac{2}{3}$ are the ratios of the matrix 
elements for the considered two-baryon states and the
 corresponding symmetric or antisymmetric two-nucleon state:
\bea
	&&	1=\frac{\matelsym{\frac{3}{2};\pm m
}{\vec T_i \cdot \vec \tau_j} }{\matelsym{1;1}{\vec\tau_i
\cdot\vec\tau_j}_{NN}} \, , \quad m=\frac{1}{2},\frac{3}{2}\, ,
 \nnl
	&&	\frac{2}{3}=\frac{\matelsym{\frac{1}{2};\pm
 \frac{1}{2}}{\vec T_i \cdot \vec \tau_j} }{\matelsym{0;0}
{\vec\tau_i\cdot\vec\tau_j}_{NN}} \,  . 
	\label{eq:isoscl2}
\eea

The scaling factors for the other isospin dependent components of the
 interaction are found in the same way.
The quark model scaling factors for all interaction components are
 listed in 
Table~\ref{tab:scl_sigppn}. As discussed for the two-charm hyperon
 states 
the results for the binding energies will be identical for states
 with different 
isospin-z projections. The same scaling factors are used and the
 small mass 
differences between the $\sigpp$ and $\sign$ are ignored here.

In Table~\ref{tab:sign_nomc} the calculated binding energies
are listed. Only the Nijm93
and the AV1' potentials suggest the existence
of very weakly bound isospin 3/2 states.
Two of the Nijmegen potentials and some of the AVn'
potentials suggest the existence of bound isospin
1/2 states. The NijmI potential predicts a much 
higher binding energy than the other interactions. For the 
Nijm93 and the AV18 potentials no bound states are found.
It should be noted that --- as for the $\Sigma_c-\Sigma_c$ 
systems --- the NijmII and the AVn' potentials prefer the 
existence of bound states with low total isospin.

\begin{table}
\begin{center}
\begin{tabular}{c|cc}
\hline\hline
	& $\wf{3/2, m}$
			& $\wf{1/2, m}$	\\
\hline
\small{Potential} 	
	& \multicolumn{2}{c}{\small{Binding Energy [MeV]}}	\\
\hline\hline
Nijm93 & $-5.3$	& ---		\\
NijmI  & ---	& $-227.3$		\\
NijmII & --- 	& $-16.9$	\\
\hline
AV18   & ---	& ---		\\
AV8'   & ---	& ---		\\
AV6'	 & ---	& ---		\\
AV4'	 & ---	& $-4.9$		\\
AVX'	 & ---	& $-1.9$ 	\\
AV2'	 & ---	& $-4.3$ 	\\
AV1'	 & $-0.2$	& $-0.2$	\\
\hline\hline
\end{tabular}
\end{center}
\caption{\label{tab:sign_nomc} Calculated binding
energies for $\Sigma_c-N$  states as obtained
    with quark model scalings.}

\end{table}

\section{Discussion}
\label{final}

The investigation above considered the use of
well known phenomenological nucleon-nucleon
interaction models to approximately describe the
interaction between nucleons and heavy flavor
hyperons and to explore the possible existence of
deuteron-like bound states of nucleons and heavy flavor
hyperons. The method relies on quark model scaling factors
for scaling the strengths of the different interaction
components to the appropriate number of light flavor
quarks in the heavy flavor hyperons considered. While
straightforward in execution the method proved to
be only of qualitative value because the large
variation in the short range components of the different
modern nucleon-nucleon interactions led to considerable
scatter in the calculated binding energies of the
two-baryon systems considered.

Another limitation is the fact that the quark model
scaling factors for the angular momentum  dependent
interaction components, which relate to the light
flavor quarks, do not reveal the expected dependence
on the hyperon mass, which is suggested by the large
$N_c$ limit of QCD \cite{kaplan,riska}. Because of this
the binding energy estimates that are obtained 
with the Nijmegen and the AVn interaction models should be
viewed as more reliable than those obtained with 
the AV18 interaction, which has a strong quadratic
angular momentum interaction.

With these provisos, the present results suggest that
nucleons form bound states with $\Xi_{c}^{'}$ hyperons
in the isospin 0 state that have binding energies
in the range between 3 and 14 MeV. With somewhat
less confidence nucleons are expected to form bound
states with $\Xi_{cc}$ hyperons, with binding
energies between 1 and 34 MeV. Such deuteron-like bound 
states of $\Xi_{cc}$ hyperons, and of $\Xi_{bb}$ as
well as of $\Xi_{cc}$ and $\Xi_{bb}$ hyperons are
also very likely, with binding energies in the range
90 - 120 MeV. Bound states of nucleons and 
$\Sigma_{c}$ hyperons are also likely, although
their binding energy cannot be estimated with 
much confidence by the present method. Most likely
states with low total isospin will be most strongly bound.

Here the interactions
between strange and light flavor quarks
have been neglected. A more realistic 
description of the states that contain strange quarks,
e.g. the $\Xi$ or $\Xi_c^{'}$ hyperons, should be possible
by using a rescaled baryon-baryon potential like the
Nijmegen NSC97 potential \cite{nsc97b}. However, due 
to the limited experimental data that is available to 
fit such potentials, this would also introduce
additional uncertainties.

\section*{Acknowledgments}

Research supported in part by the Academy of Finland through
grant 54038 and the European Euridice network
HPRN-CT-2002-00311. The work of F. F. was supported by DFG
and the European Graduate School ``Complex Systems of Hadrons
and Nuclei'' (Copenhagen-Giessen-Helsinki-Jyv\"askyl\"a).

\newpage

\begin{appendix}

\section{ Quark model wave functions}

The quark model wave functions of the baryons 
considered in this work can be written as:
\bea
	\Psi=\psi_s(r)[\phi(\mathrm{flavor})
		\chi(\mathrm{spin})]_s
		\xi_a(\mathrm{color}) \,,
\eea
where the subscript $s$ denotes a symmetric and the subscript
$a$ an anti-symmetric component of the wave function.

Nucleons and $\Xi_{(cc)}$, $\Xi^{'}_c$, $\Sigma_c$ hyperons have spin and 
flavor wave functions with mixed symmetry. The symmetric 
spin-flavor part of the total wave function is constructed as follows:
\bea
	[\phi\chi]_s=\frac{1}{\sqrt{2}}  \label{symwf}
	\left[\Pms\Cms+\Pma\Cma \right]\, .
\eea
The explicit forms of spin and flavor wave functions may be 
constructed by the methods in ref.~\cite{close}:
\bea
	\wfs{\uparrow}_{ms}=\frac{1}{\sqrt{6}}\wfs{\spupms} \, ,&&
\quad	\wfs{\uparrow}_{ma}=\frac{1}{\sqrt{2}}\wfs{\spupma}\, , \nnl
	\wfs{\downarrow}_{ms}=-\frac{1}{\sqrt{6}}\wfs{\spdnms}\; , &&
\quad	\wfs{\downarrow}_{ma}=\frac{1}{\sqrt{2}}\wfs{\spdnma} \, ,\nnl
&&
\eea
\bea
	\wf{p}_{ms}=\frac{1}{\sqrt{6}}\wf{\protms}\, , \quad &&
	\wf{p}_{ma}=\frac{1}{\sqrt{2}}\wf{\protma}\, , \nnl
	\wf{n}_{ms}=-\frac{1}{\sqrt{6}}\wf{\neutms}\ , \quad &&
	\wf{n}_{ma}=\frac{1}{\sqrt{2}}\wf{\neutma} \, ,\nnl
&&
\eea
\bea
	\wf{\Xi_{cc}^{++}}_{ms}=-\frac{1}{\sqrt{6}}\wf{\ccums}\, ,&&
\quad \wf{\Xi_{cc}^{++}}_{ma}=\frac{1}{\sqrt{2}}\wf{\ccuma}\, , \nnl
	\wf{\Xi_{cc}^{+}}_{ms}=-\frac{1}{\sqrt{6}}\wf{\ccdms} \, ,&&
\quad \wf{\Xi_{cc}^{+}}_{ma}=\frac{1}{\sqrt{2}}\wf{\ccdma}\, , \nnl
&&
\eea
\bea
	\wf{{\Xi_{c}^{'}}^{+}}_{ms}&=&\frac{1}{\sqrt{12}}\wf{\uscms}
\, , \nnl
	\wf{{\Xi_{c}^{'}}^{+}}_{ma}&=&\frac{1}{2}\wf{\uscma}\, , \nnl
	\wf{{\Xi_{c}^{'}}^{0}}_{ms}&=&\frac{1}{\sqrt{12}}
\wf{\dscms}\, , \nnl
	\wf{{\Xi_{c}^{'}}^{0}}_{ma}&=&\frac{1}{2}\wf{\dscma}\, , 
\eea
\bea
	\wf{\Sigma_c^{+}}_{ms}&=&\frac{1}{\sqrt{12}}\wf{\scpms} 
\, ,\nnl
	\wf{\Sigma_c^{+}}_{ma}&=&\frac{1}{2}\wf{\scpma}\, , \nnl
	\wf{\Lambda_c^{+}}_{ms}&=&\frac{1}{2}\wf{\lcpms}\, , \nnl
	\wf{\Lambda_c^{+}}_{ma}&=&\frac{1}{\sqrt{12}}\wf{\lcpma}
\, ,
\eea
\bea
	\wf{\Sigma_c^{0}}_{ms}=\frac{1}{\sqrt{6}}\wf{\scnms}\,,\quad&& 
	\wf{\Sigma_c^{0}}_{ma}=\frac{1}{\sqrt{2}}\wf{\scnma}\, , \nnl
	\wf{\Sigma_c^{++}}_{ms}=\frac{1}{\sqrt{6}}\wf{\scppms}\, 
,\quad &&
	\wf{\Sigma_c^{++}}_{ma}=\frac{1}{\sqrt{2}}\wf{\scppma} \, .\nnl
&&
\eea

\end{appendix}

\end{document}